# ROUGE 2.0: Updated and Improved Measures for Evaluation of Summarization Tasks

Kavita Ganesan

*Evaluation of summarization tasks is extremely crucial to determining the quality of machine generated summaries. Over the last decade, ROUGE has become the standard automatic evaluation measure for evaluating summarization tasks. While ROUGE has been shown to be effective in capturing n-gram overlap between system and human composed summaries, there are several limitations with the existing ROUGE measures in terms of capturing synonymous concepts and coverage of topics. Thus, often times ROUGE scores do not reflect the true quality of summaries and prevents multi-faceted evaluation of summaries (i.e. by topics, by overall content coverage and etc). In this paper, we introduce ROUGE 2.0, which has several updated measures of ROUGE: ROUGE-N+Synonyms, ROUGE-Topic, ROUGE-Topic+Synonyms, ROUGE-TopicUniq and ROUGE-TopicUniq+Synonyms; all of which are improvements over the core ROUGE measures.*

**1. Problems with the current ROUGE measures**

ROUGE, or Recall-Oriented Understudy for Gisting Evaluation is a method to automatically determine the quality of a summary by comparing it to another set of (ideal) summaries often created by humans (Lin and Hovy 2003)(Lin 2004a). The measure is computed by counting the number of overlapping words or n-grams between the system-generated summary to be evaluated and the ideal summaries. ROUGE by default is more of a recall oriented measure.

While ROUGE has been shown to be effective in capturing n-gram overlap between system and human composed summaries, the problem with the existing measures is that it does not give a definitive understanding of the performance of summaries in comparison to human summaries. For example, a ROUGE-1 recall score of 0.30 simply says that 30% of the content in the reference summary has been captured by the system summary. While this seems like a really low number, this score does not take into consideration synonymous concepts. The system summary could actually be effective just not capturing the exact words in the ideal summaries. In addition, if the system summary is fairly to the point but accurate and the reference summary is verbose, then the 30% that was captured could have been significant content. There currently is no way to know this because ROUGE does not allow evaluation of specific type of content coverage (i.e. topics). For example, if we compared only the topics within the ideal summaries and the system summaries, then it would become clearer if essential content has been captured.

The problem of ROUGE is much less critical if we were comparing multiple summarization systems that are solving the same task (as in TAC tasks (Dang 2005; Dang and Owczarzak 2008)). In such a case, we are primarily looking at the relative improvement of summarization systems over a baseline method. In TAC tasks, the summarization systems are typically ranked based on how much improvement is observed over the baseline method. However, when developing a summarization system in a practical setting or when developing a very new type of summarizer to solve a novel task, it is impossible to say how well the summarizer is doing just by looking at ROUGE scores.

We will now show a concrete example of two system generated summaries with corresponding reference summaries and point out problems with the ROUGE scores using this example. Example 1.1 is an example user review summary for a smart phone. The first system summary,

*SysSum1* is a very concise summary about the device and the second summary, *SysSum2* is a more verbose summary both meant to summarize the user reviews on the same smart phone. The reference summary, *RefSum* as can be seen is neither verbose nor overly concise.

**Example 1.1.**
**System Summary 1 (SysSum1):**
Lightweight phone.
Bright screen.
Screen is very clear.

**System Summary 2 (SysSum2):**
I really love this phone it is just superb, it is extremely lightweight.
Hmmm, this was actually a gift to my girlfriend and I do feel that the screen is quite nice and extremely bright.
In terms of screen, the screen is really clear and crisp.

**Reference Summary (RefSum):**
The phone is very lightweight.
The display is also very bright and clear.

**Table 1**
ROUGE-1 scores for Example 1.1

| System Summary | ROUGE-N | Recall | Precision | F-Score |
|---|---|---|---|---|
| **SysSum1** | ROUGE-1 | 0.462 | 0.750 | 0.571 |
|  | ROUGE-1+StopWordRemoval | 0.800 | 0.667 | 0.727 |
| **SysSum2** | ROUGE-1 | 0.692 | 0.196 | 0.305 |
|  | ROUGE-1+StopWordRemoval | 0.800 | 0.174 | 0.286 |

Table 1 shows the resulting ROUGE-1 scores for both *SysSum1* and *SysSum2* with and without any stop word removal. The ROUGE-1 F-Score for *SysSum1* with stop words applied is 0.727 and without stop words it is 0.571. The ROUGE-1 F-Score for *SysSum2* with stop words is 0.286 and without stop words it is 0.305.

Based on Example 1.1 and Table 1, notice that while both system summaries capture the main points of the reference summary, this is not immediately obvious from the ROUGE-1 F-Scores even with stop words removed (which should improve agreement). In the case of *SysSum2*, by looking at the F-Score, it almost appears that the system summary is of poor quality. This low F1-Score is actually caused by the low precision score due to having additional content such as '*Hmmm, this was actually a gift to my girlfriend*' even though this summary actually captures essential content from the reference summary. In fact, since *SysSum1* is very concise and captures all essential content, intuitively the ROUGE-1 F-Scores should be almost perfect. However, as can be seen the *ROUGE-1+StopWordRemoval* F-Score is only 0.727. There are several reasons as to why the ROUGE scores do not reflect content coverage accurately:

**ROUGE does not capture synonymous concepts**. ROUGE compares n-gram overlap of words on a surface level. With this, synonymous terms are not captured because the current ROUGE implementation does not provide support for synonyms. For example, the fact that the word

'display' in the reference summary above and the word 'screen' in *SysSum1* actually mean the same thing is not captured by the ROUGE-1 scores and these two words end up being treated as two different words. This yields scores that are lower than what it should be. This problem can be reduced by allowing synonyms to be captured during ROUGE scoring.

**ROUGE expects system summaries to be identical to reference summaries**. ROUGE scoring expects system summaries to exactly recover the contents of the reference summaries. Unless the system summary is completely identical to the reference summary, the ROUGE scores remain low as can be seen with both *SysSum1* and *SysSum2*. Identical summaries are rare in reality as there are different ways to express the same essential content and different set of connective words and intensifiers may be used to express the same thing. For example, *'The screen is really clear'* can very well be expressed as *'The phone display is extremely clear'* where in this case there are only 3 overlapping words out of a total of 8 unique words. This makes the agreement look artificially low. This problem can be reduced by allowing synonym capture as well as allowing systems to evaluate topic coverage as opposed to overall content coverage.

**ROUGE scores do not capture topic or subset coverage**. The current implementation of ROUGE has been focused on a complete set of n-gram overlap between reference summaries and system summaries. However, in most summarization systems it is critical to also know if different subset of content or topics have been correctly covered by system summaries. For example, in summarizing news articles, topics might be all the noun phrases in the references summaries. In summarizing tweets, topics could be just the adjectives. Another example is in opinion summarization, where topics would typically be the nouns and the adjectives (opinions). The definition of topics as can be seen is dependent on the application and the ability to analyze topic coverage would allow for optimization of the right aspects of summarization algorithms. For example, in news summarization, let us assume that the topics are defined to be all the nouns and verbs and the *ROUGE-TopicNN|VB* (coverage of noun phrases and verbs) recall is 0.90, but the *ROUGE-TopicNN|VB* precision is 0.35. This tells us that the system summaries are correctly capturing desired topics from the reference summaries but the system summaries are also including too many additional non-relevant topics. One can then optimize the summarization algorithm to pick sentences or generate abstracts that yield a more balanced *ROUGE-TopicNN|VB* precision and recall scores. Without this knowledge about topic coverage, our natural tendency would be to force the summarization algorithms to produce summaries that are identical to the reference summaries which is (a) much harder to enforce and (b) may not generalize well to new documents to be summarized.

Given the problems with ROUGE outlined in this Section, we thus propose ROUGE 2.0 which provides updated measures to address some of the outlined problems. We propose the following new measures building on the existing ones:

1.  ROUGE-{N | Topic | TopicUniq}+Synonyms - capture synonyms using a synonym dictionary (synonym dictionary customizable by application and domain)
2.  ROUGE-Topic - topic or subset coverage (topic customizable by POS occurrence)
3.  ROUGE-TopicUniq- unique topic or subset coverage (topic customizable by POS occurrence)

In Section 2.1, *Rouge-{NN | Topic | TopicUniq}+Synonyms* is introduced allowing for semantically similar words to be treated as one. In Section 2.2, *Rouge-{Topic | TopicUniq}* is introduced which allows for scoring of specific topics or subsets. Then, in Section 2.3 the Java implementation

of ROUGE 2.0 is briefly discussed. Documentation on where to download the package and how to use it is described in the following website: `http://www.rxnlp.com/rouge-2.0`.

## 2. ROUGE 2.0

ROUGE 2.0 is a Java implementation of ROUGE with improved and updated scoring. It allows capturing of semantic overlap through the use of a synonym dictionary and it also allows for evaluation of specific topics or subset of content.

### 2.1 Semantics Capture using Synonyms: Rouge-{NN | Topic | TopicUniq}+Synonyms

Rouge-{NN | Topic | TopicUniq}+Synonyms attempts to improve n-gram overlap agreement between reference summaries and system summaries by leveraging a synonym dictionary. Even though a word in the reference summary *does not* overlap with a word in the system summary on the surface, the words could in fact be synonymous. Differences in word usage are bound to happen in any language and if not accounted for, reflects poorly on the resulting ROUGE scores. As shown in Example 1.1 and Table 1, although the term 'display' and 'screen' essentially mean the same thing, these words are treated as two separate words since the current version of ROUGE only performs surface level overlap. The *ROUGE-1 + StopWordRemoval* recall scores for both *SysSum1* and *SysSum2* are 0.800 instead of 1.000 due to this surface level overlap. With the use of Synonyms (*ROUGE-1 + StopWordRemoval + Synonyms*) as shown in Table 2, notice that the recall scores for both *SysSum1* and *SysSum2* are now 1.000, clearly indicating a perfect overlap. The term 'display' and 'screen' which were previously treated as two separate words, are now considered equivalent.

In the default implementation of ROUGE 2.0, WordNet (Fellbaum 1998) is used to obtain synonyms for *nouns*, *verbs* and *adjectives* in the English language. To obtain the noun synonyms, the original *synset*, *hyponyms* and *hypernyms* with a tag count greater than 3 was used. To obtain synonyms for verbs, the *troponyms*, *hypernymes* and the original *synset* with a tag count greater than 3 was used. For the adjectives, both the adjective synset and satellite adjective synset were used as synonyms for a given adjective word.

The ROUGE 2.0 implementation is very modular in that this synonym dictionary can be replaced by any domain or language specific synonym dictionaries. For example, in the Twitter domain, there may be words that are unique to Twitter and one may choose to use a Twitter synonym dictionary for a Tweet summarization task. One can also incorporate language specific dictionaries. The format of these dictionaries can be found at `http://www.rxnlp.com/rouge-2.0`.

### 2.2 Topic or Subset Coverage: ROUGE Topic

ROUGE Topic provides the ability to evaluate different dimensions (i.e topics) of a summary. For example, in a news summarization task one may choose to evaluate coverage of all entities. These entities can be considered to be all the nouns in the reference summaries. Similarly, in evaluating opinion coverage one may consider all the nouns and the adjectives to be the topics. To support the ability to evaluate different dimensions of a summary, ROUGE-Topic allows users to specify which Part of Speech (POS) combinations should be used for evaluation. For example, *ROUGE-TopicNN | JJ* evaluates the coverage of nouns and adjectives; *ROUGE-TopicVB* evaluates the coverage of all types of verbs. These POS tags are based on the Stanford's POS Tagger (Toutanova et al. 2003) which has support for multiple languages. Table 3 shows a subset of part-of-speech tag options for ROUGE-Topic scoring.

**Table 2**
ROUGE-1 scores for Example 1.1 with the use Synonyms in ROUGE scoring.

|   |         | **ROUGE Scoring Type**                  | **Recall** | **Precision** | **F-Score** |
|---|---------|-----------------------------------------|--------|-----------|---------|
| 1 |         | ROUGE-1                                 | 0.462  | 0.750     | 0.571   |
| 2 | **SysSum1** | ROUGE-1 + Synonyms                      | 0.538  | 0.875     | **0.667** |
| 3 |         | ROUGE-1 + StopWordRemoval               | 0.800  | 0.667     | 0.727   |
| 4 |         | ROUGE-1 + StopWordRemoval + Synonyms    | 1.000  | 0.833     | **0.909** |
|   |         | **ROUGE Scoring Type**                  | **Recall** | **Precision** | **F-Score** |
| 5 |         | ROUGE-1                                 | 0.692  | 0.196     | 0.305   |
| 6 | **SysSum2** | ROUGE-1 + Synonyms                      | 0.769  | 0.217     | **0.339** |
| 7 |         | ROUGE-1 + StopWordRemoval               | 0.800  | 0.174     | 0.286   |
| 8 |         | ROUGE-1 + StopWordRemoval + Synonyms    | 1.000  | 0.217     | **0.357** |

**Table 3**
Part-of-speech options for ROUGE-Topic and ROUGE-TopicUniq scoring. Note that multiple POS options can be used concurrently and this is only a subset of POS tags that can be used. Any POS tag supported by the Stanford's POS tagger may be specified.

| **POS Options** | **Description** |
|---|---|
| **JJ**  | All types of adjectives |
| **VB**  | All verbs |
| **NN**  | All nouns including proper nouns |
| **VBD** | Verbs in past tense form |
| **RB**  | Adverbs |
| **NNP** | Proper Nouns |

ROUGE-Topic by default uses unigrams since the order of words or their co-occurrence is of less importance than the occurrence of individual topical words. Also, the dimensions evaluated can be defined based on the summarization use case. One can also choose to evaluate multiple dimensions separately. Let $REF_{i_{pos}}$ be all unigram tokens from a reference summary $i$ with the POS, $pos$. Then let $SYS_{j_{pos}}$ be all unigram tokens from a system summary $j$ with the same POS, $pos$. ROUGE-Topic can thus be computed as follows:

$$ROUGE - Topic_{recall} = \frac{\sum Overlap(REF_{i_{pos}}, SYS_{j_{pos}})}{|REF_{i_{pos}}|}$$

$$ROUGE - Topic_{precision} = \frac{\sum Overlap(REF_{i_{pos}}, SYS_{j_{pos}})}{|SYS_{j_{pos}}|}$$

where $\sum Overlap(REF_{i_{pos}}, SYS_{j_{pos}})$ represents the total number of overlapping tokens between system summary $i$ and reference summary $j$ for the selected POS, $pos$. In many cases, topic words can repeat either in the reference summaries or in the system summaries. For instance, in Example 1.1, the word 'phone' repeats 3 times in *SysSum2*. This would thus lower the *ROUGE-Topic* precision scores if only one match of 'phone' was found. Similarly, if such repetition happened in the reference summaries, this would lower the *ROUGE-Topic* recall scores. Thus, to get a more accurate understanding of how well topics are actually covered, one can use

*ROUGE-TopicUniq* where instead of counting the overlap between reference summaries and system summaries, we count the set intersection between the two. Let $REFUniq_{i_{pos}}$ be a unique set of unigram tokens from a reference summary *i* with the POS, *pos*. Then let $SYSUniq_{j_{pos}}$ be unique unigram tokens from a system summary *j* with the same POS, *pos*. *ROUGE-TopicUniq* can thus be computed as follows:

$$ROUGE - TopicUniq_{recall} = \frac{REFUniq_{i_{pos}} \cap SYSUniq_{j_{pos}}}{|REFUniq_{i_{pos}}|}$$

$$ROUGE - TopicUniq_{precision} = \frac{REFUniq_{i_{pos}} \cap SYSUniq_{j_{pos}}}{|SYSUniq_{j_{pos}}|}$$

Table 4 shows the resulting *ROUGE-Topic* and *ROUGE-TopicUniq* scores for Example 1.1 where topics in this case are considered to be nouns (NN) and adjectives (JJ). The intuition for selecting nouns and adjectives in this example is to allow evaluation of opinion coverage. Based on Table 4, just by analyzing the *ROUGE-TopicNN|JJ* recall scores (rows 1 and 5), we get a sense that most of the opinions in the reference summary have been correctly captured by both *SysSum1* and *SysSum2*. These scores further improve with the use of synonyms (+Synonyms, rows 2 and 6). From the *ROUGE-TopicUniqNN|JJ* scores (rows 3 and 7) we can see that agreement further improves over *ROUGE-TopicNN|JJ* in terms of precision[1]. This shows that repetition is being suppressed and we are only accounting for topic matches once. The *ROUGE-TopicUniqNN|JJ + Synonyms* F-Score for *SysSum1* shows that the system summary recovers all the opinions from the reference summaries and also the system summary is concise with no unnecessary topics and opinions in the mix. This is in contrast to the *ROUGE-TopicUniqNN|JJ + Synonyms* scores for *SysSum2* where the precision score is still much lower than the recall. This shows that there are additional topics and opinions in the mix such as 'screen is quite nice' and 'this phone it is just superb'. With this, we know that the summarization algorithm is picking up all the key topics, but the type of sentences that are being used or generated contain additional unnecessary information. Given this knowledge, we can focus on tuning the summarization algorithm to select sentences or generate abstracts that are less verbose.

### 2.3 ROUGE 2.0 Package Implementation

While the original ROUGE package was developed in Perl (Lin 2004b), the ROUGE 2.0 package with updated measures has been developed in Java since there have been many issues with getting the Perl version working with Windows and many Mac and Linux machines. Also, the Java version simplifies the entire *system* and *reference* summary naming convention and formatting allowing researchers to focus on evaluation. Since this package is platform independent, it would enable a broader outreach to all researchers and industry developers. The complete ROUGE 2.0 package along with source code and documentation can be accessed from this website address: `www.rxnlp.com/rouge-2.0`

To verify that the output of ROUGE 2.0 is in fact accurate, a sanity check was done to verify that the original ROUGE-N scores from the Perl package were identical to ROUGE-N scores from the Java implementation. It turns out that for ROUGE-N, where N > 1, there seems to be '1' added to all counts in the Perl implementation (Lin and Hovy 2003). It is not clear if this is a smoothing effect. Since this is not mentioned in the formula for computation of ROUGE scores this 'add 1' feature will thus not be replicated in ROUGE 2.0. Code submissions for updated and improved scoring mechanisms that have been peer evaluated are encouraged.

---

[1] In this particular example, only precision improves but in other cases no improvement may be observed or both recall and (or) precision may improve

**Table 4**
ROUGE-TopicNN|JJ and ROUGE-TopicUniqNN|JJ scores based on Example 1.1

|   |         | RougeTopic                              | Recall | Precision | F-Score |
|---|---------|-----------------------------------------|--------|-----------|---------|
| 1 |         | ROUGE-TopicNN\|JJ                       | 0.800  | 0.667     | 0.727   |
| 2 | SysSum1 | ROUGE-TopicNN\|JJ + Synonyms            | **1.000** | **0.833** | **0.909** |
| 3 |         | ROUGE-TopicUniqNN\|JJ                   | 0.800  | 0.800     | 0.800   |
| 4 |         | ROUGE-TopicUniqNN\|JJ + Synonyms        | **1.000** | **1.000** | **1.000** |
|   |         | **RougeTopic**                          | **Recall** | **Precision** | **F-Score** |
| 5 |         | ROUGE-TopicNN\|JJ                       | 0.800  | 0.308     | 0.444   |
| 6 | SysSum2 | ROUGE-TopicNN\|JJ + Synonyms            | **1.000** | **0.385** | **0.556** |
| 7 |         | ROUGE-TopicUniqNN\|JJ                   | 0.800  | 0.364     | 0.500   |
| 8 |         | ROUGE-TopicUniqNN\|JJ + Synonyms        | **1.000** | **0.455** | **0.625** |